\documentclass[a4paper,11pt]{article}
\usepackage{pos}
\usepackage{enumitem}
\usepackage{lineno}

\title{Predictions for gamma-rays from clouds associated with supernova remnant PeVatrons}
 \ShortTitle{Predictions for $\gamma$-rays from clouds associated with SNR PeVatrons}

\author*[a]{Alison Mitchell}
\author[b]{Gavin Rowell}
\author[c,d]{Silvia Celli}
\author[b]{Sabrina Einecke}

\affiliation[a]{Department of Physics, ETH Zurich,\\
  CH-8093 Zurich, Switzerland}

\affiliation[b]{University of Adelaide, School of Physical Sciences,\\
Adelaide, SA 5005, Australia}

\affiliation[c]{Dipartimento di Fisica dell’Universita` La Sapienza,\\
P. le Aldo Moro 2, I-00185 Rome, Italy}
\affiliation[d]{Istituto Nazionale di Fisica Nucleare, Sezione di Roma,\\
P. le Aldo Moro 2, I-00185 Rome, Italy}

\emailAdd{amitchell@phys.ethz.ch}

\abstract{Interstellar clouds can act as target material for hadronic cosmic rays; gamma-rays produced through inelastic proton-proton collisions and spatially associated with the clouds provide a key indicator of efficient particle acceleration.
However, even for PeVatron sources reaching PeV energies, the system of cloud and accelerator must fulfil several conditions in order to produce a detectable gamma-ray flux.
In this contribution, we characterise the necessary properties of both cloud and accelerator.
Using available Supernova Remnant (SNR) and interstellar cloud catalogues, and assuming particle acceleration to PeV energies in a nearby SNR, we produce a ranked shortlist of the most promising target systems; those for which a detectable gamma-ray flux is predicted.
We discuss detection prospects for future facilities including CTA and SWGO; and compare our predictions with known gamma-ray sources, including the Ultra-High-Energy sources recently detected by LHAASO.
A range of model scenarios are tested, including variation in the diffusion coefficient and particle spectrum, under which the best candidate clouds in our shortlist are consistently bright.
On average, a detectable gamma-ray flux is more likely for more massive clouds; for systems with lower separation distance between the SNR and cloud; and for slightly older SNRs, due to the time required for particles to traverse the separation distance.}

\FullConference{37$^{\rm{th}}$ International Cosmic Ray Conference (ICRC 2021)\\
		July 12th -- 23rd, 2021\\
		Online -- Berlin, Germany}

\begin{document}
\maketitle

\section{Introduction}

Interstellar clouds in the vicinity of cosmic particle accelerators provide a means to trace the origins of energetic Galactic Cosmic Rays (CRs). Supernova remnants are among the most promising candidate source classes for the acceleration of cosmic rays to PeV energies (i.e. `PeVatron' behaviour). Particles accelerated to the highest energies will escape the accelerator region at early times in its evolution, entering the interstellar medium (ISM). 
Interstellar clouds, therefore, provide a suitable target material for the cosmic rays to undergo hadronic interactions, leading to the production of a gamma-ray signal.  

Experimental facilities such as Imaging Atmospheric Cherenkov Telescope (IACT) arrays and ground-based particle detectors (such as HAWC and LHAASO) have in recent years begun to probe the gamma-ray sky above $>100$\,TeV, providing the first evidence for Galactic PeVatrons. 
In this contribution, we summarise the key points from our recent modelling study \cite{Mitchell21} and discuss these in the context of the PeVatron sources recently reported by LHAASO \cite{lhaasopev}. 

\noindent Our model describes:
\begin{enumerate}[label=\roman*)]
    \item the energy-dependent release time of particles from an SNR with expanding radius  $R_{\mathrm{esc}} = R_{\mathrm{SNR}}(t=t_{\mathrm{esc}}(E))$,
    \item the impulsive injection of particles with a power-law spectrum $J_p(E,R,t)= N_0\cdot E^{-\alpha}\cdot f(E,R,t)$, with probability distribution function $f(E,R,t)$ from \cite{AA96},
    \item particle diffusion through the ISM and the cloud (see equation \eqref{eq:diffcoeff}), 
    \item the resulting gamma-ray flux due to particle interactions at the cloud from \cite{Kelner06}.
\end{enumerate}  

\noindent The momentum dependence of the escape time $t_{\mathrm{esc}}(p)$ is described by: 
\begin{equation}
   t_{\mathrm{esc}}(p) = t_{\mathrm{sed}}\left(\frac{p}{p_M}\right)^{-1/\beta}~\mathrm{yr,}
   \label{eq:tesc}
\end{equation}
where $\beta=2.5$ is assumed, and $p_M$ is the maximum momentum achieved by particles at the Sedov time ($t_{\mathrm{sed}}$), set to $1\,$PeV c$^{-1}$ for PeVatrons \cite{Gabici07,Celli19}. A Sedov time of $t_{\mathrm{sed}}\sim 1.6$\,kyr suitable for core collapse supernovae is adopted as default. 

\begin{figure}
    \centering
    \includegraphics[width=0.48\textwidth]{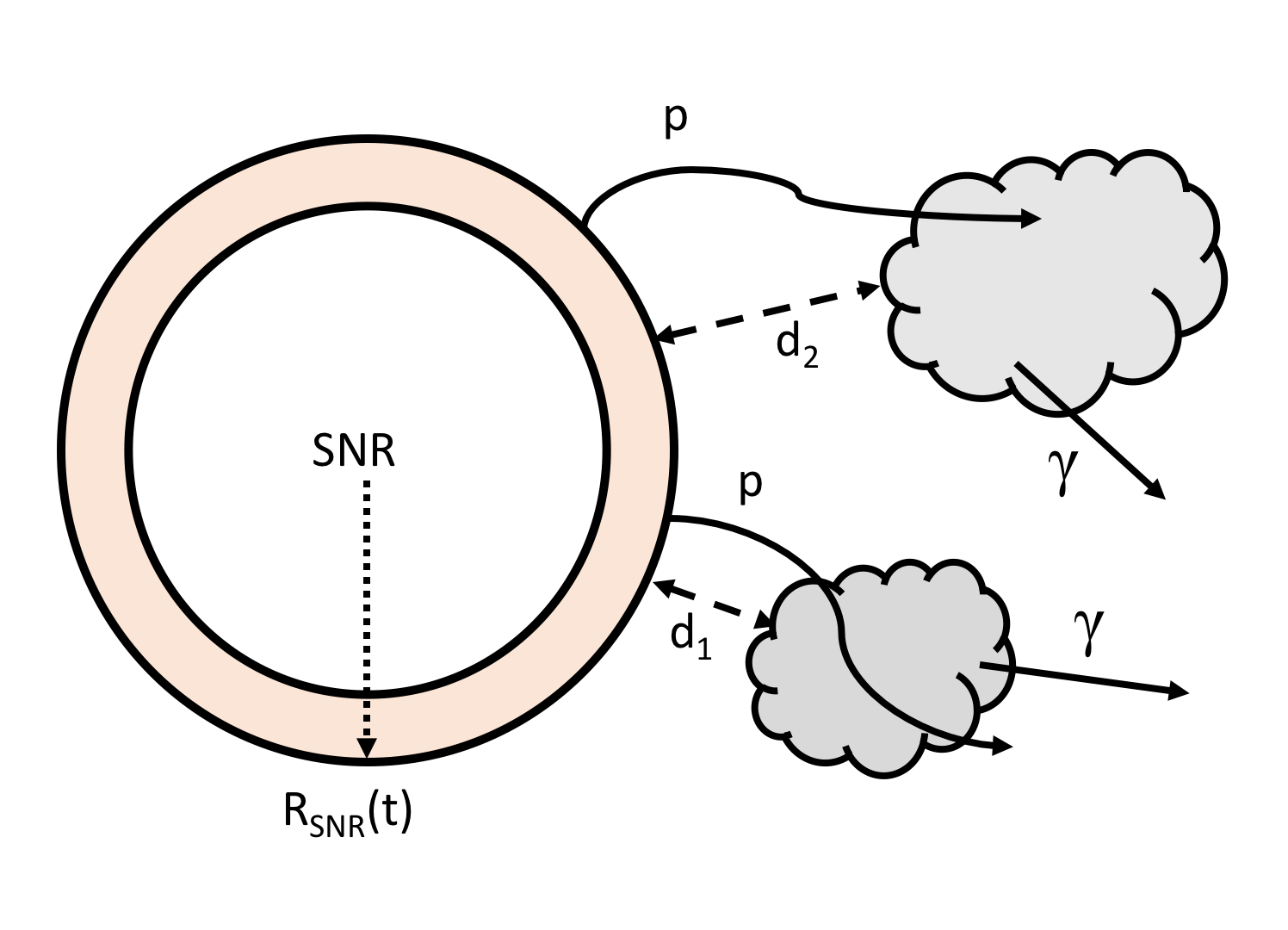}
    \caption{The geometry assumed in this model. Particles escape from the expanding SNR radius $R_{\mathrm{SNR}}(t)$ and travel through the ISM to nearby clouds, at distances $d_1$ and $d_2$ from the SNR. Interactions of the particles (assumed to be dominated by protons) with the cloud produce gamma-rays.}
    \label{fig:schematic}
\end{figure}

\noindent The energy dependence of the diffusion coefficient is described by: 
\begin{equation}
D(E) = \chi D_0 \left(\frac{E (\mathrm{GeV}) }{B(n)/3\mathrm{\mu G}}\right)^{\delta}~\mathrm{cm^2/s},
\label{eq:diffcoeff}
\end{equation}
with $\delta=0.3$ and the magnetic field is assumed dependent on the density only for $(n \geq 300 \mathrm{cm}^{-3})$, taking a value of $10\mu$G otherwise. Note also that the diffusion suppression coefficient $\chi$ is assumed to be 1 in the ISM and 0.05 otherwise. 
Figure \ref{fig:schematic} illustrates the general case considered by our model to predict the gamma-ray flux from interstellar clouds. In section \ref{sec:modelresults} we show some of the key dependencies of the gamma-ray flux on parameters of the cloud-SNR system of CR target and accelerator.  

In order to make predictions for real systems, we searched for suitable pairs of clouds and SNRs in close proximity to each other ($\Delta d \lesssim 200$\,pc) based on the catalogue of clouds from \cite{Rice16} and the catalogue of SNRs from \cite{GreenSNR}. For cases where the true SNR distance was unknown, we assumed that the SNR was located at the same distance along the line of sight as that of the cloud. Where the SNR age was unknown, this was estimated from the measured SNR radius by inverting the expression for the expansion of the SNR over time (and assuming Sedov phase evolution). 

\section{Model Results}
\label{sec:modelresults}

\begin{figure}
    \centering
    \includegraphics[width=0.5\textwidth]{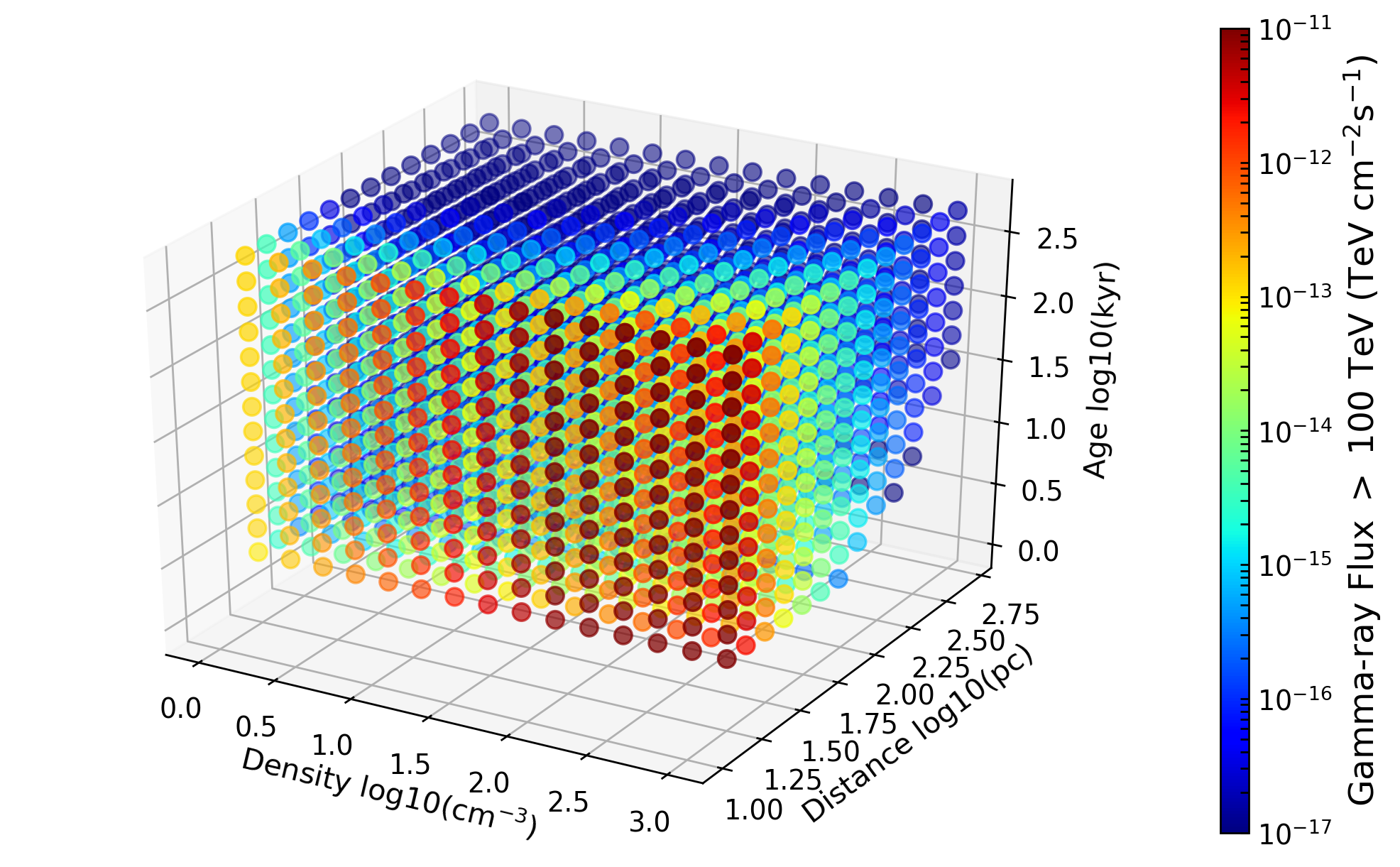}
    \includegraphics[width=0.48\textwidth]{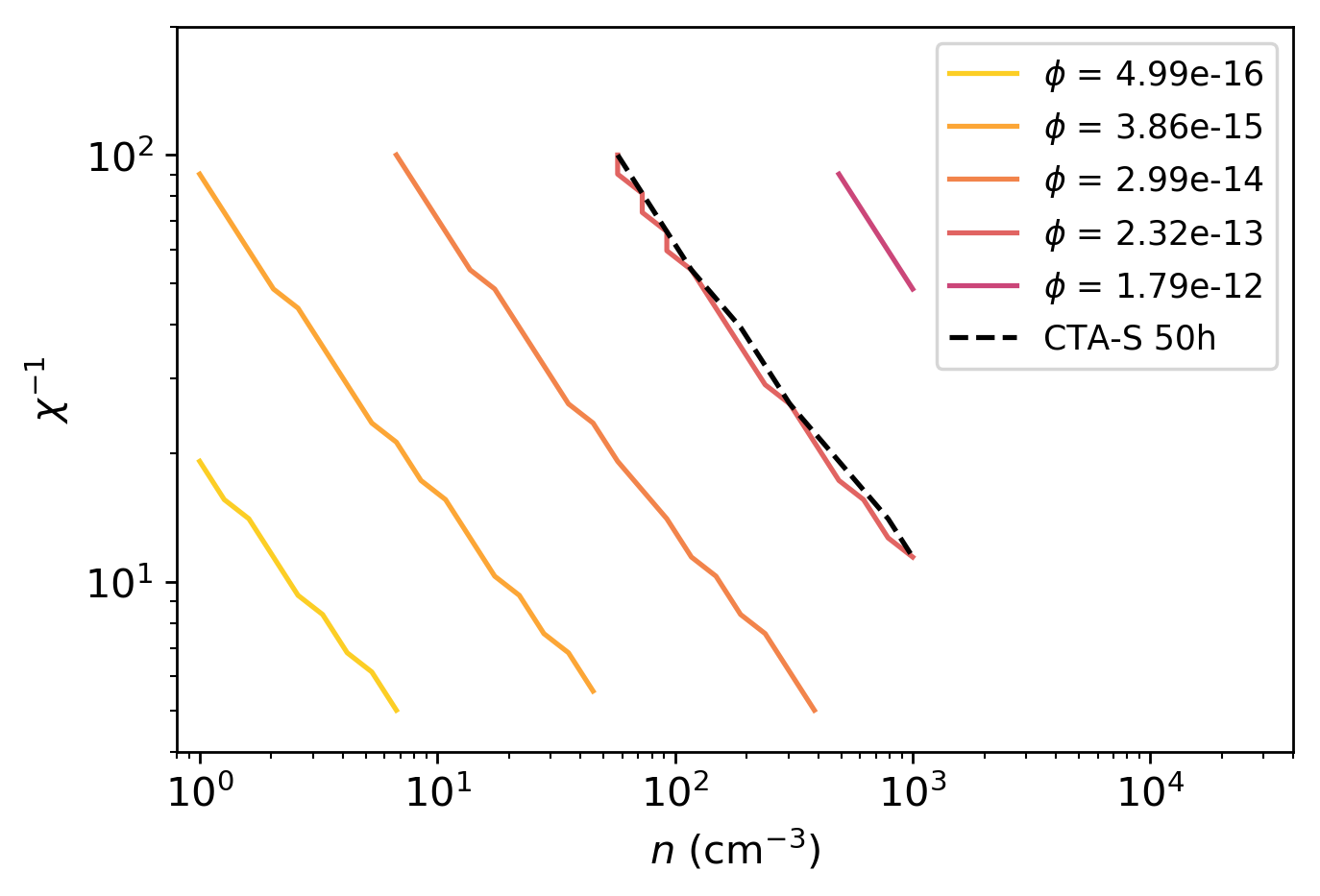}
    \caption{Predictions for the integral $\gamma$-ray flux $>100$\,TeV as a function of the SNR-cloud system properties. Left: Variation with density of the cloud, separation distance between the cloud and the SNR, and the age of the SNR. Right: Variation with density of the cloud and $\chi$, the diffusion suppression coefficient. The cloud radius is fixed to 10\,pc and the distance of the system from Earth is fixed to 1\,kpc.  }
    \label{fig:phasespace}
\end{figure}

Figure \ref{fig:phasespace} illustrates the phase space of the gamma-ray flux dependence on system properties. The 3D plot in the left hand panel shows that the gamma-ray flux increases with increasing cloud density, expected due to the increase in target material available for hadronic interactions. Similarly, the gamma-ray flux decreases with increasing distance, as fewer of the energetic particles are able to reach the target cloud. An interesting feature to note, however, is that at large separation distances the predicted gamma-ray flux increases with increasing age; i.e. older accelerators are preferred, simply due to the time it takes particles to travel through the ISM to reach the cloud. 

The right hand side of figure \ref{fig:phasespace} shows the only plot where the value of $\chi$ is varied. Depending on the diffusion, the same gamma-ray flux level can be achieved for different cloud densities, with a higher gamma-ray flux reached for a more suppressed diffusion (larger $\chi^{-1}$). This is because a so-called slow diffusion, with lower $D(E)$, is more efficient at filling the cloud, rather than simply traversing. Hence, the CRs remain in the vicinity of the cloud for longer and encounter more target material for interactions. 

\begin{figure}
    \centering
    \includegraphics[width=0.5\textwidth]{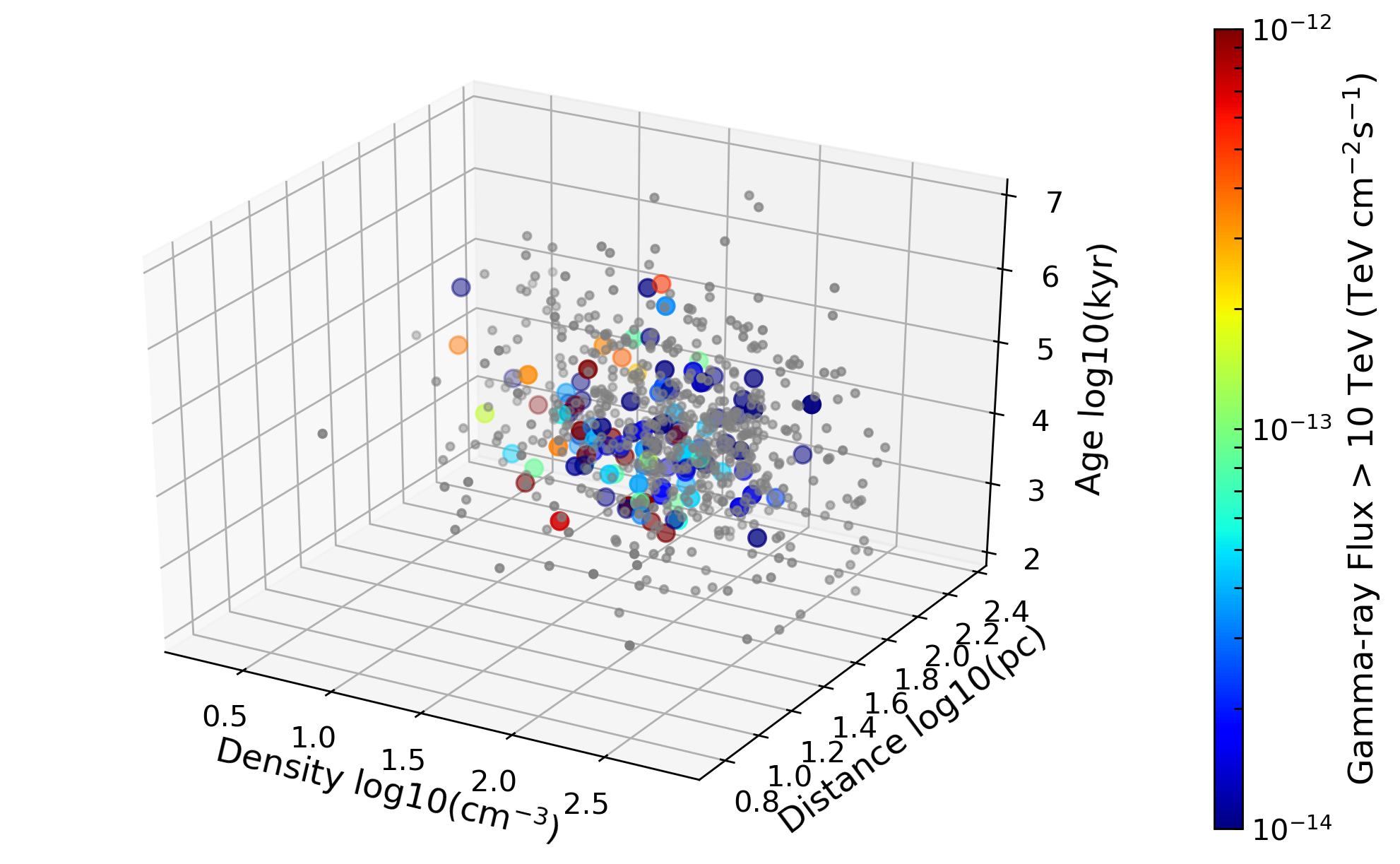}
    \includegraphics[width=0.48\textwidth]{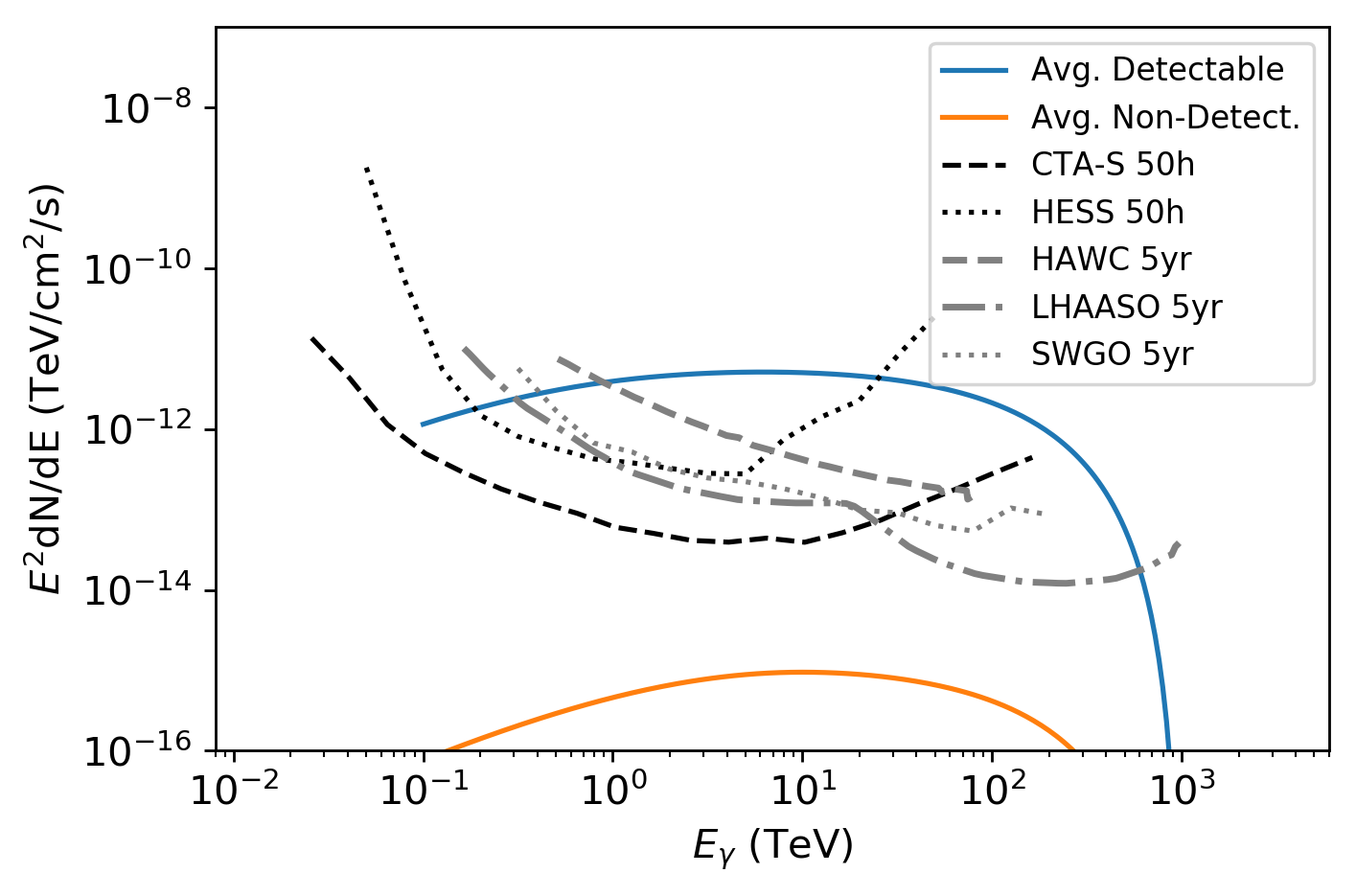}
    \caption{Predictions of the integral $\gamma$-ray flux $>100$\,TeV for pairs of SNRs and clouds. Left: as a function of the cloud density, separation distance and SNR age phase space. Coloured points are detectable systems; grey points are non-detectable. Right: $\gamma$-ray spectra for the average properties of detectable and non-detectable systems, compared to the sensitivity of current and future $\gamma$-ray facilities. }
    \label{fig:gfluxclouds}
\end{figure}

The distribution of real SNR-cloud systems in this phase space is shown in the left hand panel of figure \ref{fig:gfluxclouds}. It is clear that although density, separation distance and SNR age are primary factors governing the predicted gamma-ray flux, they are not the only factors, with the distance of the system from Earth (among others) also playing a key role. Grey dots show systems for which the predicted flux is too low to be considered detectable by current or future gamma-ray facilities. 
In the right hand side of figure \ref{fig:gfluxclouds}, we show the gamma-ray spectrum obtained from SNR-cloud systems with the average properties of (in blue) systems with a detectable flux and (in orange) systems with a non-detectable flux. For comparison, the sensitivity curves of current and future facilities are shown, where LHAASO performs impressively compared to other facilities in the key energy range $>100$\,TeV. 

On average, detectable systems were found to:
\begin{enumerate}[label=\roman*)]
    \item have larger clouds (more target material),
    \item be located closer to the SNR (a shorter distance for particles to travel to reach the cloud), and
    \item be associated with older SNRs, as it takes time for the particles to traverse the ISM. 
\end{enumerate}

As several model assumptions were made, we varied several of these to investigate which of the predictions were most robust, with consistently high fluxes predicted under several model scenarios. These included: fast and slow values of the diffusion coefficient ($3\times10^{27}{\rm cm}^{2}{\rm s}^{-1}$ and $3\times10^{26}{\rm cm}^{2}{\rm s}^{-1}$ at 1\,GeV respectively, fast value as default); varying the index of the power law injection spectrum $\alpha$ (1.8, 2.2 and adopting 2.0 as default); and exchanging the expression for the Sedov time from one appropriate for core collapse supernovae (1.6\,kyr) to one suitable for type Ia supernovae (234\,yr).    
We note that core collapse (or type II) supernovae were assumed to be the default case, as these occur when a star reaches the end of nuclear burning in its red giant stage, and are more likely to be found in rich molecular environments. By contrast, the type Ia supernovae that occur due to mass accretion in a binary system from a massive companion onto a white dwarf are associated to older systems and hence environments that are less rich in molecular interstellar material.  

\section{Discussion of Selected Candidate Targets}
\label{sec:targets}

Table \ref{tab:bestclouds} summarises the most promising cloud candidates identified by our study \cite{Mitchell21}. These were selected as some of the brightest predicted fluxes, with more reliable predictions (as a known distance for the associated SNR was available) and with a detectable flux consistently predicted across various model scenarios. 

Three of these clouds are located outside the coverage of the H.E.S.S. Galactic Plane Survey (HGPS) \cite{HGPS}, hence no upper limits $F_\gamma^{UL}$ from that dataset were derived. Systematic uncertainties due to model assumptions were estimated as being at least one order of magnitude. This is sufficient to explain most of the discrepancies between the flux upper limit $F_\gamma^{UL}$ and the predicted flux $F_\gamma^{\mathrm{total}}$, yet the brightest cloud candidate (1) at (330.050, 0.130) stands out. This cloud is predicted to be powered by CRs escaped from the SNR G337.2-00.7. Although the angular distance between the two appears rather large, this corresponds to a distance of just a few hundred parsec that can be traversed by CRs within the available time.\footnote{The available time is evaluated as the SNR age less the escape time for particles of a given energy.} 
Although this is feasible with the diffusion coefficient adopted, it is possible that we have been optimistic with our assumption that CRs can propagate over such a large distance through the ISM between accelerator and cloud essentially unimpeded. A similar argument likely accounts for the lack of gamma-ray emission coincident with the cloud (10) at (322.600, -0.630).

\begin{table}[]
    \centering
    \begin{tabular}{c|l|l|l|l|l|l}
	\hline
	 ID & Cloud coordinates & Size & Distance  & $F^{UL}_\gamma >$ 100\,TeV & $F^{\mathrm{total}}_\gamma$ $>$ 100\,TeV  & SNR \\
	  & (l, b) deg & deg & kpc & TeV cm$^{-2}$ s$^{-1}$ & TeV cm$^{-2}$ s$^{-1}$ & \\
	 \hline
    1 & (330.050, 0.130) & 0.398 & 3.970 & 1.03e-12 & 4.90e-09 &  G337.2-00.7 \\ 
    6 & (84.780, -0.410) & 0.689 & 5.840 & N/A & 6.23e-12  & G084.2-00.8  \\ 
7 & (32.700, -0.210) & 0.172 & 4.810 & 9.41e-13 & 3.23e-12 &  G032.8-00.1\\ 
9 & (339.340, -0.410) & 0.299 & 5.770 & 1.26e-12 & 2.93e-12 &  G344.7-00.1 \\ 
10 & (322.600, -0.630) & 0.242 & 3.450 & 3.65e-13 & 2.74e-12 & G330.2+01.0 \\ 
11 & (36.100, -0.140) & 0.492 & 3.400 & 6.14e-12 & 1.52e-12 & G035.6-00.4 \\ 
12 & (332.960, -0.040) & 0.177 & 10.850 & 1.60e-12 & 1.30e-12 & G337.8-00.1 \\ 
14 & (114.730, 1.100) & 0.639 & 0.840 & N/A & 7.20e-13 & G114.3+00.3 \\ 
17 & (113.610, -0.670) & 0.965 & 2.620 & N/A & 3.67e-13 &  G113.0+00.2 
    \end{tabular}
    \caption{Summary of the flux predictions from the most promising candidate clouds; those for which the distance to both SNR and cloud are constrained. Flux upper limits were taken from the HGPS \cite{HGPS}. }
    \label{tab:bestclouds}
\end{table}

\begin{figure}
    \centering
    \includegraphics[width=1.05\textwidth,height=3.5cm]{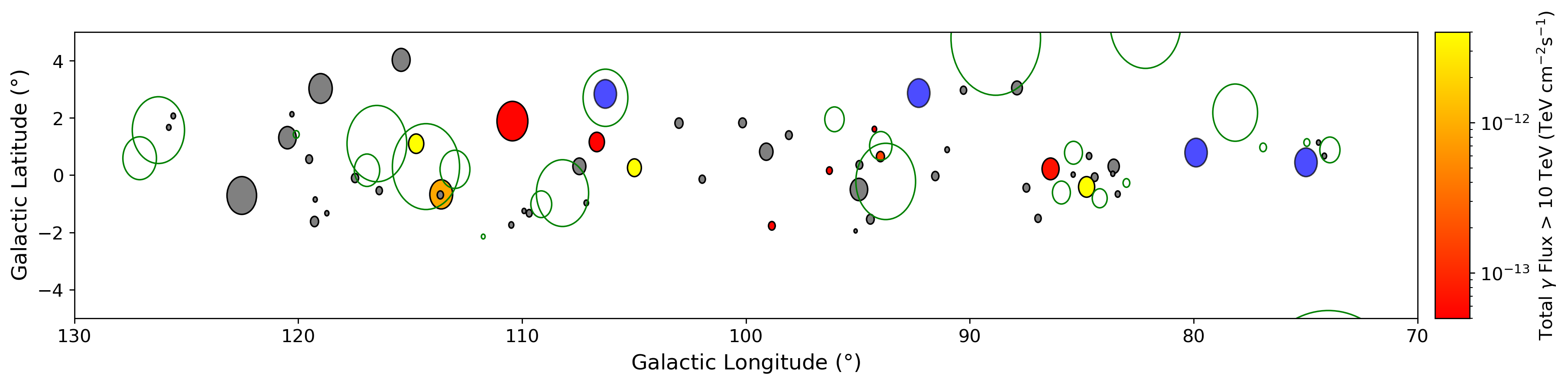}
    \caption{Region of the Galactic plane from $130^\circ > l > 70^\circ$ not covered by the HGPS. Locations of LHAASO sources $>100$\,TeV are shown in blue; locations of SNRs are indicated by green circles. Clouds with a $\gamma$-ray flux prediction from this model are shown in colour scale or grey for a non-detectable flux prediction. }
    \label{fig:GPnorth}
\end{figure}
\begin{figure}
    \centering
    \includegraphics[trim=3cm 0mm 0mm 0mm, clip,width=1.1\textwidth]{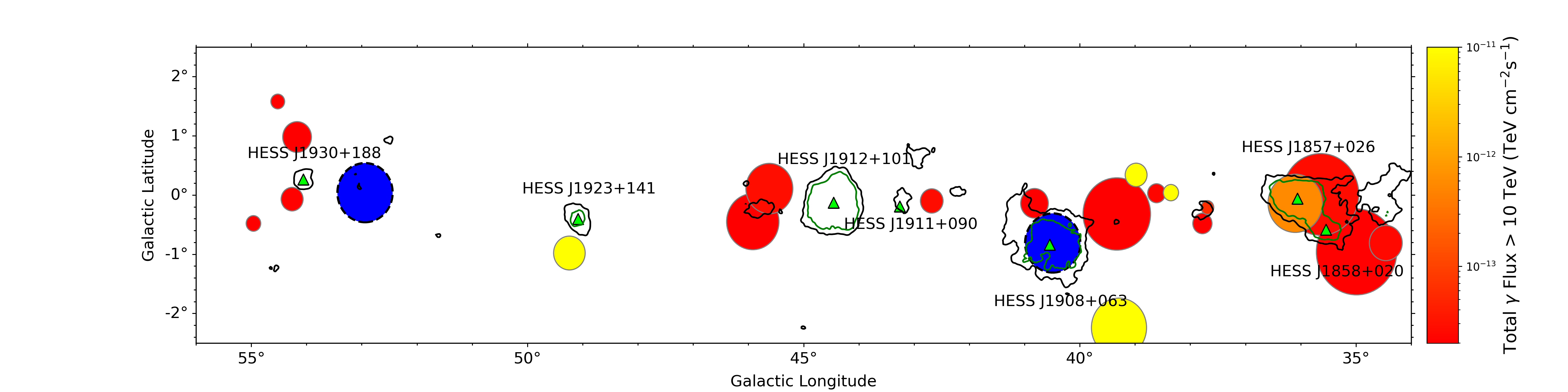}
    \includegraphics[trim=3cm 0mm 0mm 0mm, clip,width=1.1\textwidth]{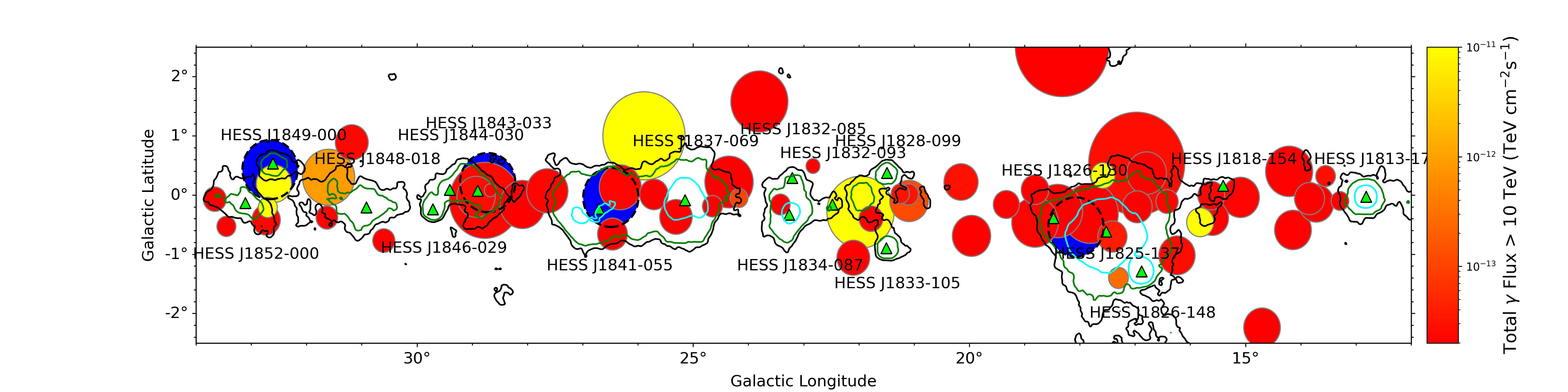}
    \caption{Region of the Galactic Plane covered by both H.E.S.S. and LHAASO, with the locations of LHAASO sources $>100$\,TeV shown in blue and $\gamma$-ray flux predictions for clouds shown colour scale. The locations of H.E.S.S. sources are indicated by green triangles and by significance contours. Note that not all of the predicted fluxes are considered reliable predictions, see text and \cite{Mitchell21} for details. Top panel: $56^\circ > l > 34^\circ$; Bottom panel: $34^\circ > l > 12^\circ$. }
    \label{fig:galplanehess}
\end{figure}

The three targets from table \ref{tab:bestclouds} that are not covered by HGPS, due to their more northerly location along the Galactic plane, are shown in figure \ref{fig:GPnorth}. Dedicated observations may be necessary to further constrain the flux predictions in this region, although we note that no coincident emission has yet been reported by the sky-scanning facilities HAWC and LHAASO. The sources detected by LHAASO at the highest energies \cite{lhaasopev} are indicated in figures \ref{fig:GPnorth} and \ref{fig:galplanehess} in blue. Although the nature of these sources is not yet firmly established, the majority can be positionally associated to known energetic pulsars. 

This leaves four clouds in table \ref{tab:bestclouds}; two of these are not show in either figure \ref{fig:GPnorth} or \ref{fig:galplanehess} of this proceeding; cloud 9 at (339.340, -0.410) and cloud 12 at (332.960, -0.040) are coincident with two known TeV gamma-ray sources, HESS\,J1646-458 (or Westerlund\,1) and HESS\,J1616-508 respectively. In both cases, the predicted emission is compatible with the derived upper limit, implying that it will be challenging to disentangle a gamma-ray flux signal from these clouds from the surrounding emission. 

The remaining two clouds are shown in figure \ref{fig:galplanehess}; cloud 11 at (36.100, -0.140) is coincident with the source HESS\,J1857+026, of unidentified origin (see also the discussion in section 5.1.3 of \cite{Mitchell21}). Cloud 7, at (32.700, -0.210) is immediately adjacent to LHAASO\,J1849-0003, one of the recently identified PeVatron sources that could plausibly be associated with either the likely pulsar wind nebula HESS\,J1849-000 or with the young massive cluster W43 \cite{lhaasopev}.  

It should be noted that all bar one of the LHAASO sources shown in figure \ref{fig:galplanehess} are coincident with known H.E.S.S. sources; generally associated with energetic pulsars. LHAASO\,J1929+1745 (at $l=52.9^\circ$ is, however, also coincident with a known TeV source detected by HAWC; 2HWC\,J1928+177, also associated with an energetic pulsar. 

Figure \ref{fig:galplanehess} shows clouds that are predicted by the model to have an integral flux $>10$\,TeV detectable by the Cherenkov Telescope Array (CTA). A much larger number of clouds for which the predicted flux is below the detection threshold are not shown. Conversely, additional predictions beyond table \ref{tab:bestclouds} from \cite{Mitchell21}, such as 
flux contributions from hypothetical SNRs (as traced by pulsars) are included in the total flux depicted in figure \ref{fig:galplanehess}. In table \ref{tab:lhaaso_cloud} we provide a brief summary of our model predictions for clouds that are coincident with LHAASO sources from \cite{lhaasopev}. For most cases, the flux predictions are far below the fluxes measured by LHAASO above 100\,TeV, with the exception of J1843-0338, where the SNR G028.6-00.1 is a plausible source of the emission in both cases. 

For further discussion, please refer to the full study \cite{Mitchell21}.

\begin{table}[]
    \centering
    \begin{tabular}{c|l|l|l|l|l}
     Cloud coordinates & Size & Distance  & $F^{\mathrm{total}}_\gamma$ $>$ 100\,TeV  & SNR & LHAASO \\
	  (l, b) deg & deg & kpc & TeV cm$^{-2}$ s$^{-1}$ & & \\
	 \hline
         (17.81,-0.31) & 0.3458 & 6.3 & 2.14e-16 & G017.0-00.0 & J1825-1326 \\
         (26.69,-0.07) & 0.2018 & 4.01 & 6.96e-17 & G027.8+00.6 & J1839-0545 \\
         (28.67,0.08) & 0.1400 & 6.77 & 1.51e-14 & G028.6-00.1 & J1843-0338 \\
         (32.59,0.59) & 0.1722 & 6.23 & 1.09e-17 & G031.9+00.0 & J1849-0003
    \end{tabular}
    \caption{Model predictions for clouds coincident with the location of LHAASO sources. }
    \label{tab:lhaaso_cloud}
\end{table}

%
%
%

\end{document}